\documentclass[doublecol]{epl2} 

\usepackage[utf8]{inputenc}
\usepackage{ulem}
\usepackage{color,graphicx}
\usepackage{amssymb,amsmath}
\usepackage{bm,bbm}
\usepackage{verbatim}
\usepackage{float,soul,mathtools}
\usepackage{physics}
\usepackage{bbold,mathrsfs}
\usepackage{lineno}
\usepackage{xcolor}
\usepackage{hyperref}
\usepackage{breakurl}

\newcommand{\be}{\begin{eqnarray}}
\newcommand{\ee}{\end{eqnarray}}
\newcommand{\beq}{\begin{equation}}
\newcommand{\eeq}{\end{equation}}

\newcommand{\mbb}{\mathbbm}
\newcommand{\mf}{\mathfrak}
\newcommand{\mr}{\mathrm}

\title{Emergence of classical realism under successive noncommuting measurements}

\author{D. M. Fucci\inst{1} \and L. F. Gaissler\inst{1} \and R. M. Angelo\inst{1}}
\shortauthor{Fucci \etal}

\institute{                    
  \inst{1} Department of Physics, Federal University of Paran\'a, P. O. Box 19044, 81531-980 Curitiba, PR, Brazil.
}

\abstract{
The problem of emergence of classicality from quantum mechanics has been addressed over time through numerous frameworks, from Bohr's correspondence principle to quantum Darwinism. Traditional approaches associate the emergence of classicality with the decoherence process induced by large reservoirs on the system's state. In this work, we present an effective mechanism by which classicality emerges through the establishment of elements of reality. This involves the process of successive monitoring of noncommuting observables. To assess physical reality, we employ the realism criterion introduced by Bilobran and Angelo [EPL, 112 (2015) 40005], as well as their quantifier for the violations of this criterion. With these tools, we formally demonstrate, for generic systems, that a quasi-classical regime can always be reached with a sufficiently large number of incompatible measurements. Thus, instead of diagnosing the emergence of the classical regime in terms of the resulting algebraic characteristics for the density operator under the action of large reservoirs, our results reveal that classicality can emerge, at the level of physical elements of reality, from the coupling of the system with environments of a few degrees of freedom.}

\begin{document}

\maketitle

\section{Introduction}

Experiments conducted at the turn of the 19th century revealed that classical physics required a profound revision. This led to the birth of quantum mechanics, a theory built upon an algebraic structure radically different from that of classical physics. The conceptual and mathematical differences between these theories established the perception that the new theory has serious foundational issues such as the measurement problem~\cite{Brukner2017,Wheeler2014}, interpretational ambiguities~\cite{Jammer1974}, or nonlocality~\cite{Bell1964,Brunner2014}. 

However, it may very well be the case that the old physics was merely a provisional view of nature, valid only in specific regimes, such as those involving heavy and ideally isolated systems. This hypothesis, which tacitly assumes that quantum mechanics is universal and thus capable of recovering classical mechanics in some limit, became known as the problem of the quantum-to-classical transition. Achieving a better understanding of this transition is crucial for consolidating the universality of quantum mechanics and for addressing the alleged foundational problems, elevating them to the level of natural behaviors. This work aims to contribute to this discussion.

The investigation of the quantum-to-classical transition began as soon as the formal foundations of quantum mechanics were proposed. From an instrumentalist stance, Bohr approached this issue arguing that quantum mechanics descriptions only make sense given an instrumental setup. According to this view, quantum states core feature, complementarity, becomes then necessarily tied to the intrinsic limitations of measurement protocols~\cite{Bohr1935}. Ehrenfest, in turn, pursued a different approach by, without referring to measurements, demonstrating how the expectation values of observables for large objects approximate classical behavior~\cite{Ehrenfest1927}. Recent times have witnessed the development of the decoherence program~\cite{Zurek1982,Zurek2003,Schlosshauer2007,Schlosshauer2021}, in which aspects of classicality are derived from the coupling with (and subsequent neglect of) infinite, unobservable degrees of freedom, generically referred to as the environment. Zurek's quantum Darwinism~\cite{Zurek2009-QD,Chen2019,Unden2019,Zurek2023} underpins to a great extent the \textit{status quo}. It conceives the environment as responsible for decohering the quantum state and redundantly encoding its properties in pointer states. As a result, multiple observers can independently access environmental fragments and reach a consensus on the properties of the system, thus establishing a notion of objectivity~\cite{Zurek2003}.

There is, though, an aspect that is central to all the aforementioned approaches: the establishment of realism, understood as value definiteness in the system's physical quantity. Firstly introduced in the literature by Einstein, Podolsky and Rosen~\cite{Einstein1935}, the notion of elements of reality was associated with scenarios where ideal predictability is achievable without the implication of any disturbance on the system. Today there is not a unique notion of realism or physical reality (see Ref.~\cite{Dieguez2018} for a brief review of the subject), but it is still reasonable to consider this concept as an indicator of classicality. More specifically, within this framework, the emergence of classical reality occurs when all physical quantities assume definite values (even if not known by the observer).

Among the available frameworks for physical realism, we choose the criterion devised by Bilobran and Angelo (BA)~\cite{Bilobran2015}, as it possesses operational meaning and includes a quantifier for its violation. Axiomatized in \cite{Orthey2022}, this criterion allowed for novel notions of nonlocality~\cite{Gomes2018, Orthey2019, Fucci2019}, and advancements in resource theory~\cite{Costa2020} and quantum foundations~\cite{Dieguez2018, Engelbert2020, Paiva2023}, backed by experimental tests~\cite{Mancino2018,Dieguez2022}. A distinctive feature of this realism criterion is that it allows us to diagnose the presence of elements of reality not only based on the quantum state, which is an algebraic quantity with debatable interpretation, but mainly through analyses of how the information associated with the state is disturbed by measurements. It is precisely this characteristic that allows the criterion to be stated in a theory-independent manner~\cite{fucci2024}, giving it solid and general foundations.

Our main result is built on a mechanism of sequential nonselective measurements in succession on observables $A$ and $B$. This procedure implements a simplified model of a system under continuous environmental monitoring of at least two distinct observables. In this view, each nonselective measurement represents an entangling interaction between the system and an environmental degree of freedom, followed by the subsequent tracing out (discarding) of that environmental component. Continuous monitoring thus corresponds to a repeated cycle of interaction and discard, effectively implementing a sequence of measurements on the system without the need for infinitely many environmental degrees of freedom. We show that this process alone is sufficient for a realism-based notion of classicality to emerge.

The upcoming section provides a brief introduction on BA's criterion and aspects of the generalized Bloch sphere formalism, necessary for the presentation of our main result. Then, we present a bound for the lack of realism of the observables upon pairwise sequential measurements of incompatible observables. Following that, we apply it for a qubit case and, taking advantage of the simplicity of qubits and qutrits, we give a more strict bound, illustrated by numerical analysis.

\section{Preliminaries} 

BA's approach to realism postulates that $A$ is an element of reality whenever a nonselective projective measurement of this observable, mathematically represented as
\begin{equation} \label{def: phi}
\Phi_A(\rho) \coloneqq \sum_{i} A_i \rho A_i = \sum_i p_{a_i} A_i,
\end{equation}
is unable to change the original state $\rho$. Here, $p_{a_i} = \operatorname{Tr} [A_i \rho]$ is the probability associated with the measurement outcome $a_i$ and $A_i$ is the corresponding projector. Formally, BA's criterion prescribes that $A$ is an element of reality for $\rho$ \textit{iff}
\begin{equation} \label{eq: BA crit}
    \rho = \Phi_A(\rho).
\end{equation}

To quantify deviations in their criterion, the authors introduced the so-called \textit{irreality}. Let $S(\rho) \coloneqq -\operatorname{Tr}[\rho \ln{\rho}]$ be the von Neumann entropy and $S(\rho||\sigma) \coloneqq \operatorname{Tr}[\rho (\ln{\rho} - \ln{\sigma})]$ the relative entropy. The irreality of $A$ given $\rho$ is expressed by:
\begin{equation}\label{def: irr}
    \mathfrak{I}_A(\rho) \coloneqq S\big(\rho||\Phi_A(\rho)\big) = S\big(\Phi_A(\rho)\big) - S(\rho).
\end{equation}
Regarding single-partite states, the map $\Phi_A$ decoheres the state by an amount quantified by $\mf{I}_A(\rho)$.

Parallels between emergence of realism and quantum Darwinism are seen when considering the Stinespring's theorem~\cite{Spehner2014} connection with the map $\Phi_A$. It is synthesized in the equality
\begin{equation}\label{eq: stinespring}
    \Phi_A(\rho)=\operatorname{Tr}_{\mathcal{E}}\left[U\left(\rho \otimes\left|e_0\right\rangle\left\langle e_0\right|\right) U^{\dagger}\right],
\end{equation}
being a unitary operator $U$ acting on $\mathcal{H} \otimes \mathcal{H}_\mathcal{E}$ and $\ket{e_0} \in \mathcal{H}_\mathcal{E}$. This relation allows the understanding of the nonselective measurement as being conducted by an environment $\mathcal{E}$. Described by the state $\ket{e_0}\bra{e_0}$, it stores the information about $\rho$ in a dynamic process and is then discarded. In this way, both frameworks achieve decoherence through the interaction between the system and the environment, followed by the discard of the environmental degrees of freedom.

It is worth mentioning that, as originally conceived for bipartite systems, irreality is sensitive not only to coherence but also to quantum correlations. Additionally, as reported elsewhere \cite{fucci2024}, both the criterion and the measures of its violations admit extensions to generalized probabilistic theories.

The intuition behind this realism notion may be enhanced by giving it a geometrical interpretation. To accomplish this, we represent quantum states as vectors in a Euclidean real-valued vector space, as the Bloch sphere representation of qubit states. Here, we present necessary aspects of the generalized Bloch sphere representation to address higher-dimensional quantum states. More detailed presentations of the formalism are provided in \cite{Aerts2014} and \cite{Aerts2016}.

The generators of the special group of degree $d$, SU($d$), give us a basis for representing linear operators acting on a $d$-dimensional state space. Those are given by the set $\left\{\mathbbm{1}, \Lambda_1, \ldots, \Lambda_{d^2-1}\right\}$, where $\Lambda_i$ are complex $d \times d$ self-adjoint orthogonal traceless matrices. By requiring a normalization condition $\Tr(\Lambda_i \Lambda_j)=2 \delta_{i j}$, any $d$-dimensional quantum state can be represented as
\begin{equation} \label{eq: gen rho}
    \rho_{\vec{r}} = \frac{1}{d} \left(\mathbbm{1} + C_d \vec{r} \cdot \vec{\Lambda}\right).
\end{equation}
Here, $C_d = \sqrt{d(d-1) / 2}$, and, defining an orthonormal basis $\left\{\hat{e}_i\right\}_{i=1}^{d^2-1}$ in $\mathbbm{R}^{d^2-1}$,  $\vec{r}=\sum_{i=1}^{d^2-1} r_i \hat{e}_i$ is a vector and $\vec{\Lambda}=\sum_{i=1}^{d^2-1} \Lambda_i \hat{e}_i$ is a matrix-valued vector. This enables the encoding of a density operator as a real vector $\vec{r}$ of dimension $d^2 - 1$ in a real ball $B(\mathbbm{R}^{d^2-1})$. It is easy to check that the setting $d=2$ establishes $\Lambda_i$ as the Pauli matrices, and that \eqref{eq: gen rho} reduces to the usual Bloch sphere representation.

To represent projective operators, where $\sum_i A_i = \mathbbm{1}$ and $\operatorname{Tr}\left(A_i A_j\right)=\delta_{i j}$, we write
\begin{equation} \label{eq: gen op}
    A_i = \frac{1}{d} \left(\mathbbm{1}+C_d \vec{a}_i \cdot \vec{\Lambda}\right),
\end{equation}
where $\vec{a}_i$ satisfy $\sum_i \vec{a}_i = \vec{0}$ and $\vec{a}_i \cdot \vec{a}_j = (\delta_{i j} d-1) /(d-1)$, describing the vertices of a dimension $d-1$ regular simplex, $\triangle_A$. Embedded in the hypersphere and containing its center, $\triangle_A$ gives the region populated by vectors representing all normalized convex sums of $A$ eigenvectors. Being $a_i$ the eigenvalues of a traceless observable $A = \sum_i a_i A_i$, one can express $A = \vec{a} \cdot \vec{\Lambda}$ with $\vec{a} = (C_d / d) \sum_i a_i \vec{a}_i$. The algebra determined by the generators $\Lambda_i$ together with the vector products we introduced lead to $\operatorname{Tr}[(\vec{r}_1\cdot\vec{\Lambda})(\vec{r}_2\cdot\vec{\Lambda})]=2(\vec{r}_1\cdot\vec{r}_2)$, enabling one to determine the probabilities
\begin{equation} \label{eq: gen p}
    p_{a_{i}}=\operatorname{Tr}(A_i\rho_{\vec{r}})=\frac{1}{d}[1+(d-1)\vec{a}_i\cdot\vec{r}].
\end{equation}
Lastly, the unrevealed measurement protocol map \eqref{def: phi} fits this formalism, as found in \cite{Martins2020}, according to
\begin{equation} \label{eq: gen phi}
    \Phi_A (\rho_{\vec{r}}) = \frac{1}{d} (\mathbbm{1} + C_d P_A\vec{r} \cdot \vec{\Lambda}), \quad P_A\vec{r} = \frac{d-1}{d} \sum_{i=1}^d (\vec{a}_i \cdot \vec{r}) \vec{a}_i,
\end{equation}
where $P_A(\bullet)\coloneqq \frac{d-1}{d}\sum_i \vec{a}_i\cdot(\bullet)\,\vec{a}_i$ is an idempotent, nonorthogonal projector onto the $A$ simplex. Like this, we recontextualize the condition \eqref{def: irr} such that $A$ is an element of reality for $\rho$ \textit{iff} $\vec{r}=P_A\vec{r}$, happening \textit{iff} $\vec{r} \in \triangle_A$.

\section{Main result}

Here, we demonstrate how the irreality of any observable $X$ in a given state $\rho$ diminishes through environmental monitoring. To illustrate this effect, we introduce a toy model consisting of pairwise sequential measurements of noncommuting observables performed by two agents. This scenario shares the same fundamental mathematical structure as the previously considered case, enabling a direct comparison and insight into how continuous environmental monitoring influences the emergence of classicality.

\noindent {\bf Theorem.} \textit{For any state $\rho_{\vec{r}}$, there always exists a sufficiently large number of sequential pairwise measurements of noncommuting observables $A$ and $B$ that leads the irreality $\mathfrak{I}_X$ of any observable $X$ to be sufficiently small.}

\vskip2mm
\noindent \textit{Proof}.---We begin by handling Fanne's inequality \cite{Audenaert2007}, $|S(\rho)-S(\sigma)|\leq T \ln{(d-1)} + H_{\text{bin}}(T)$, for any $d\in\mathbbm{N}_{\geq 2}$. Here, $T=T(\rho,\sigma)=\frac{1}{2}||\rho-\sigma||_1$, where $||O||_1=\operatorname{Tr}\sqrt{O^\dagger O}$ is the Schatten $1$-norm of the operator $O$, and $H_\text{bin}(T)=-T\ln{T}-(1-T)\ln{(1-T)}$ is the binary entropy. Since $T\in [0,1]$, we can utilize the inequalities $T\leq \sqrt{T}$ and $H_\text{bin}(T)\leq \sqrt{2T}$ (which can be readily checked graphically) to arrive at $|S(\rho)-S(\sigma)|\leq \left(\sqrt{2}+\ln(d-1)\right)\sqrt{T}$. Now, using Hölder's inequality\footnote{Hölder's inequality~\cite{Horn2012} is the statement that $||AB||_1\leq ||A||_p||B||_q$, where $p,q\in [1,\infty]$, $\frac{1}{p}+\frac{1}{q}=1$, and $||A||_p=\left(\Tr (A^\dag A)^{p/2} \right)^{1/p}$. With $p=q=2$ and $B=\mathbbm{1}/d$, one finds $||A||_1\leq \sqrt{d}\,||A||_2$.}, $||\rho-\sigma||_1\leq \sqrt{d}\,||\rho-\sigma||_2$, we obtain
\begin{align}
	|S(\rho)-S(\sigma)|\leq d^{1/4}\left(1+\frac{\ln(d-1)}{\sqrt{2}}\right)\sqrt{||\rho-\sigma||_2},
\end{align}
where $||O||_2=\sqrt{\Tr(O^\dag O)}$ stands for the Schatten 2-norm. In terms of the Bloch representation, direct calculation yields $||\rho_{\vec{r}_2}-\rho_{\vec{r}_1}||_2=\sqrt{(d-1)/d}\,||\vec{r}_2-\vec{r}_1||$, from which it follows that
\begin{align}\label{eq:ineq1}
	\big|S\big(\rho_{\vec{r}_2}\big)-S\big(\rho_{\vec{r}_1}\big)\big|\leq g(d)\sqrt{||\vec{r}_2-\vec{r}_1||},
\end{align}
where $g(d)\coloneqq [(d-1)]^{1/4}\left(1+\frac{\ln(d-1)}{\sqrt{2}}\right)$, a monotonically increasing function of $d$, and $||\vec{r}||=\sqrt{\vec{r}\cdot\vec{r}}$.

For the next step, we consider a nonselective measurement of a spin observable (a traceless Hermitian operator) $A=\vec{a}\cdot\vec{\Lambda}$ over the state $\rho_{\vec{r}}$, resulting in $\Phi_A(\rho_{\vec{r}})$. Sequentially, a nonselective measurement is performed of an observable $B=\vec{b}\cdot\vec{\Lambda}$, for which a similar mathematical structure holds, then yielding $\Phi_B \Phi_A(\rho_{\vec{r}}) = \frac{1}{d} \big( \mathbb{1} + C_d P_B P_A \vec{r} \cdot \vec{\Lambda} \big)$. Thus, after a sequential measurement of the pair $\{A,B\}$, the transition is represented in the Bloch formalism as $\vec{r}\to \vec{r}_1\equiv P_BP_A\vec{r}$. From the properties of projectors, it readily follows that $||\vec{r}_1||\leq ||\vec{r}||$, which allows us to write $||\vec{r}_1||=\epsilon_1||\vec{r}||$ with $\epsilon_1\in\mbb{R}_{[0,1]}$. Note that $\vec{r}_1=\vec{r}$ when $\vec{r}=\vec{0}$ (a maximally mixed state) or when $\{\rho,A,B\}$ share the same simplex. In the latter case, $\epsilon_1=1$. For $n$ sequential measurements, the state will be given by 
\begin{align}
	\vec{r}_n=(P_BP_A)^n\vec{r}
\end{align} 
with
\begin{align}\label{eq:r_n}
    ||\vec{r}_n||=\left(\Pi_{k=1}^n\epsilon_k\right)||\vec{r}||\quad \left(\forall\, n\in \mathbbm{N}_{> 0}\right),
\end{align}
where we employed the product notation $\Pi_{k=1}^n$ and introduced $\epsilon_k\in\mathbb{R}_{[0,1]}$ such that $||\vec{r}_k||=\epsilon_k||\vec{r}_{k-1}||$.

We are now able to conclude the proof. Considering the irreality $\mf{I}_X(\rho_{\vec{r}_n})=S\big(\Phi_X(\rho_{\vec{r}_n})\big)-S(\rho_{\vec{r}_n})$ of a generic spin operator $X=\vec{x}\cdot\vec{\Lambda}$, we rely on the inequality \eqref{eq:ineq1} to write
\begin{align}
	\mf{I}_X(\rho_{\vec{r}_n})&\leq g(d)\sqrt{||\vec{r}_n-P_X\vec{r}_n||}\nonumber \\
    &= g(d) \sqrt{\frac{||(1-P_X)(P_B P_A)^n \vec{r}||}{||(P_B P_A)^n \vec{r}||}} \sqrt{||(P_B P_A)^n \vec{r}||} \nonumber \\
	&= g(d)\sqrt{||(1-P_X)\hat{r}_B||}\,\sqrt{||(P_BP_A)^n\vec{r}||},\label{eq:ineq2}
\end{align}
where $\hat{r}_B\equiv\hat{r}_n=\vec{r}_n/||\vec{r}_n||$ lies in the $B$ simplex. At this point, it is opportune to realize that the dependence on the number $n$ of pairwise measurements disappears when the triple $\{\rho,A,B\}$ share the same simplex, i.e., $[A,B]=[A,\rho]=[B,\rho]=0$, as in this case, $\vec{r}_n=\vec{r}$. Also, for $\vec{r}=\vec{0}$, the equality $\mf{I}_X(\rho_{\vec{r}_n})=0$ holds for all $n$ and regardless of the choices of $A$, $B$, and $X$, highlighting the reason why $\mbb{1}/d$ is termed a classical state. When $P_A\vec{r}=0$, $A$ and $\rho$ eigenstates form mutually unbiased bases\footnote{Two observables $X$ and $Y$ are said to form mutually unbiased bases if their eigenbases $\{\ket{x_i}\}$ and $\{\ket{y_j}\}$ satisfy the property that for all $i$ and $j$, the squared modulus of the inner product between their eigenvectors is constant, i.e., $|\langle x_i|y_j\rangle|^2=1/d$, where $d$ is the dimension of the Hilbert space.} (MUB), implying that $\vec{r}$ lies on the simplex of an observable that is maximally incompatible with $A$. Under these circumstances, we find $\mf{I}_X(\rho_{\vec{r}_n})=0$ even for $n=1$, indicating that a single $A$ measurement can establish classical realism. Finally, since $\hat{r}_n$ lies on the $B$ simplex, meaning that $B$ is an element of reality for $\rho_{\vec{r}_n}$, one will also obtain $\mf{I}_X(\rho_{\vec{r}_n})=0$ whenever $[X,B]=0$. 

Under the assumption that $[A,B]\neq 0$, we now return to Eq.~\eqref{eq:r_n} resorting to the notation $\epsilon_k\leq\mr{O}(\epsilon)$ with $\mr{O}(\epsilon)=\min_{1 \leq k \leq n} \epsilon_k$. Hence, $||\vec{r}_n||/||\vec{r}||$ is limited according to $\Pi_{k=1}^n\epsilon_k\leq[\mr{O}(\epsilon)]^n$. With this, we can conclude that $\mf{I}_X(\rho_{\vec{r}_n})\leq\delta$ when $n\geq [ n_{\min} ]$, where $[\bullet]$ denotes the integer part of $\bullet$. Explicitly, using the right-hand side of Eq.~\eqref{eq:ineq2} to write
\begin{equation}
    \delta = g(d) \sqrt{||(1-P_X)\hat{r}_B||}\,\sqrt{[\mathrm{O}(\epsilon)]^{n_{min}} ||\vec{r}||}
\end{equation}
and solving it for $n_{\min}$, we get
\begin{align}\label{eq: nmin}
	n_{\min}=2\frac{\ln{\left(\frac{\delta}{g(d)\sqrt{||\vec{r}||\,||(1-P_X)\hat{r}_B||}}\right)}}{\ln{\left[\mr{O}(\epsilon)\right]}}.
\end{align}
For $\delta$ vanishingly small, this last equation guarantees that the irreality of any observable $X$ vanishes with at least $n_{\min}$ sequential pairwise measurements, as we wanted to demonstrate.\hfill{\tiny $\blacksquare$}

A remark is now in order. As is apparent already in equation \eqref{eq:r_n}, each pairwise monitoring has the effect of decreasing the modulus (hence, the purity) of the post-measurement state. Therefore, whenever $[A,B]\neq 0$, the asymptotic limit ($n\to\infty$) will always be such that $\rho_{\vec{r}_n}\to\mathbbm{1}/d$ and $\mf{I}_X(\rho_{\vec{r}_n})\to 0$. This shows that environments with infinitely many degrees of freedom are not demanded for classical reality to emerge. A continuous dynamics involving the coupling of the system with a two-degree-of-freedom environment would suffice. Also, when $A$ and $B$ form MUB, then $||\vec{r}_1||=||P_BP_A\vec{r}||=0$ and $\mf{I}_X(\rho_{\vec{r}_1})=0$. In this case, a single interaction with a bipartite environment already yields classical reality. This regime highlights the fundamental difference between our approach and quantum Darwinism~\cite{Zurek2009-QD}. In the latter, classical reality requires objectivity, which, in turn, necessitates large environments capable of redundantly encoding information and being accessed by several observers.

\section{Examples} 

In this section, we will use further mathematical results, different from those employed in the previous section, which will also yield upper bounds for irreality. Using these results, we will independently demonstrate the consistency of the theorem for the cases where $d=2$ and $d=3$; that is, we will show that, except when $[A,B]=0$, every observable will become an element of reality as $n\to\infty$.

Let us start analyzing a qubit scenario. Setting $d=2$ gives $\rho_{\vec{r}}=\frac{1}{2}(\mathbbm{1}+\vec{r}\cdot\vec{\sigma})$, where $\vec{\sigma}$ gives the Pauli matrices and a generic spin observable $A=\hat{a}\cdot\vec{\sigma}$ determines a one-dimensional simplex. Once $P_A\vec{r}=(\hat{a}\cdot\vec{r})\hat{a}$ and $P_BP_A\vec{r}=(\hat{a}\cdot\hat{b})(\hat{a}\cdot\vec{r})\hat{b}$, we find $\vec{r}_n=(\hat{a}\cdot\hat{b})^{2n-1}(\hat{a}\cdot\vec{r})\hat{b}$ and $P_X\vec{r}_n=(\hat{a}\cdot\hat{b})^{2n-1}(\hat{a}\cdot\vec{r})(\hat{b}\cdot\hat{x})\hat{x}$. We now use the fact that
\begin{equation}
    H_{\text{bin}}\left(\frac{1+\mu \lambda}{2}\right)-H_{\text{bin}}\left(\frac{1+\lambda}{2}\right)\leq \lambda^2 (1-\mu^4)\,\ln{2},
\end{equation}
where $H_{\text{bin}}(p)=-p \ln{p}-(1-p)\ln{(1-p)}$ is Shannon's binary entropy, $\lambda\in\mbb{R}_{[0,1]}$, and $\mu\in\mbb{R}_{[-1,1]}$. For $\mu=0$ and $\lambda=1$, one can verify that the above upper bound is tight. With this, we obtain
\begin{align}
\mf{I}_X(\rho_{\vec{r}_n})&=H_\text{bin}\left( \frac{1+||P_X\vec{r}_n||}{2}\right)-H_\text{bin}\left( \frac{1+||\vec{r}_n||}{2}\right)\nonumber \\
&\leq \big(\hat{a}\cdot\vec{r}\big)^2 \big(\hat{a}\cdot\hat{b}\big)^{2(2n-1)}\left[1-\big(\hat{x}\cdot\hat{b}\big)^4\right]\ln{2}.
\end{align}
Since $|\hat{a}\cdot\hat{b}|<1$ whenever $[A,B]\neq 0$, this formula corroborates our main result: we can always choose $n$ large enough to upper bound irrealism to arbitrarily small values for any  $X$. Physically, this means that continuous monitoring will inevitably drive the system to the classical realism regime.

Now we examine qutrits, as they exhibit the complexity typical of $d>2$ systems while remaining computationally friendly. Physical states are now represented on an eight-dimensional hypersphere, with embedded triangles corresponding to the two-dimensional simplices determined by observables. For instances where $d\geq 3$, we can resort to a more convenient bound:
\begin{equation} \label{def: irr max}
    \mf{I}_X(\rho_{\vec{r}})\leq \max_{\{X\}} \mathfrak{I}_X (\rho_{\vec{r}}) = \ln{d} - S(\rho_{\vec{r}}) \eqqcolon I(\rho_{\vec{r}}).
\end{equation}
That this bound is strict can be verified by taking $\vec{r}$ to lie within the simplex $\triangle_Y $ of $Y$, such that $X$ and $Y$ form MUB. In this case, $P_X\vec{r}=\vec{0}$ and $S(\rho_{P_X\vec{r}})=\ln{d}$.

\begin{figure}[t] 
\centering
\includegraphics[width=1\columnwidth]{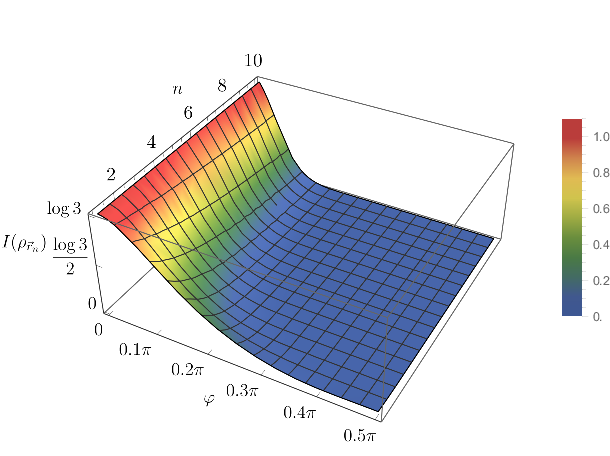} 
\caption{Plot of $I(\rho_{\vec{r}_n})$ as a function of the angle $\varphi$ between $\hat{a}$ and $\hat{b}$, and the number of successive pairs of measurements $n$ for a qutrit state.}
\label{fig: plot}
\end{figure}

A numerical analysis was conducted to determine the bound \eqref{def: irr max} of $\rho_{\vec{r}_n}$ by taking $\rho_{\vec{r}}$ corresponding to the qutrit eigenstate of $S_z$ with eigenvalue $-1$, $A=S_{\vec{z}}\,\cos{\varphi}+S_{\vec{x}}\,\sin{\varphi}$ and $B=S_{\vec{z}}$, which gives $\hat{a}\cdot\hat{b}=\cos{\varphi}$. Without the necessity of employing the generalized Bloch sphere formalism, using Eq.~\eqref{def: phi} we derive $I(\rho_{\vec{r}_n}) = \ln{3} - S[(\Phi_B\Phi_A)^n(\rho)]$ to calculate the results showcased in Figure \ref{fig: plot}. We considered $n\in\mbb{N}_{\leq10}$ and $\varphi\in[0,\pi/2]$ to illustrate the diminishing bound for $\mf{I}_X(\rho_{\vec{r}_n})$ with the increase of $\varphi$ and $n$. As $n$ increases, the bound decreases more steeply as a function of $\varphi$. This tendency reflects the fact demonstrated in Eq.~\eqref{eq: nmin} that, for a sufficiently large $n$, the bound approaches zero even when $[A,B]\approx0$,  here corresponding to small values of $\varphi$. We remark that, contrasting with the qubit case, having $\varphi=\pi/2$ does not imply a MUB. In fact, for $n=1$, we have $I(\rho_{\vec{r}_n})\approx 0.015$.

\section{Conclusions}

Taking the definiteness of all physical properties (realism) as a paradigm of classicality, we proved that successive monitoring of noncommuting observables is a sufficient mechanism for the emergence of classical behavior. Although this does not ensure objectivity in the quantum Darwinism sense, which would demand the proliferation of information in several parts of an environment, it emphasizes that some aspects of classicality do not require coupling with large environments to manifest. Specifically, we derived equation \eqref{eq: nmin}, which prescribes the order of sequential pairwise measurements for the irreality of any observable to fall within a bound, and we provided an illustration for a qubit as well as a numerical depiction for a qutrit.

Our results reinforce the idea that quantum mechanics is a universal theory that recovers classical behavior under certain conditions. Furthermore, our approach highlights the emergence of classicality with respect to ontological aspects, an approach that remained unexplored until now. In particular, we show that even low-structure reservoirs can drive physical systems toward classical reality.

We leave open for future work the question of whether pairs of positive operator-valued measures can also be effective in establishing realism-based classicality; in that case, notions more sophisticated than noncommutativity are required to diagnose measurement incompatibility.

\acknowledgments
This study was financed in part by the Coordena\c{c}\~ao de Aperfei\c{c}oamento de Pessoal de N\'ivel Superior - Brasil (CAPES) - Finance Code 001. R.M.A. thanks the financial support from the National Institute for Science and Technology of Quantum Information (CNPq, INCT-IQ 465469/2014-0) and the Brazilian funding agency CNPq under Grants No. 305957/2023-6.

\bibliography{refs}

\end{document}